\documentclass{article}
\usepackage{arxiv}

\usepackage[utf8]{inputenc} 
\usepackage[T1]{fontenc}    
\usepackage{hyperref}       
\usepackage{url}            
\usepackage{booktabs}       
\usepackage{amsfonts}       
\usepackage{nicefrac}       
\usepackage{microtype}      
\usepackage{graphicx}
\usepackage{doi}
\usepackage[utf8]{inputenc} 
\usepackage[T1]{fontenc}
\usepackage{url}
\usepackage{ifthen}
\usepackage{cite}
\usepackage{graphicx}
\usepackage{amsfonts}
\usepackage{amssymb}
\usepackage{bbm}
\usepackage{subfigure}

\title{TinyTurbo: Efficient Turbo Decoders on Edge}
\author{\textbf{S.~Ashwin Hebbar}$^{1\dag}$ \and \textbf{Rajesh Mishra}$^{2\dag}$ \and \textbf{Sravan Kumar Ankireddy}$^2$ \and \textbf{Ashok Vardhan Makkuva}$^1$ \and \textbf{Hyeji Kim}$^2$ \and \textbf{Pramod Viswanath}$^1$}
\date{} 					

\newcommand{\undertitle}{}

\usepackage{algorithmic, algorithm}
\usepackage{mathtools}

\newcommand\blfootnote[1]{%
  \begingroup
  \renewcommand\thefootnote{}\footnote{#1}%
  \addtocounter{footnote}{-1}%
  \endgroup
}

\usepackage{comment}

\usepackage{thmtools}

\usepackage{prettyref,xspace}
\usepackage{tikz}

\newrefformat{cond}{Condition~\ref{#1}}
\newrefformat{eq}{(\ref{#1})}
\newrefformat{thm}{Theorem~\ref{#1}}
\newrefformat{th}{Theorem~\ref{#1}}
\newrefformat{chap}{Chapter~\ref{#1}}
\newrefformat{sec}{Section~\ref{#1}}
\newrefformat{algo}{Algorithm~\ref{#1}}
\newrefformat{fig}{Figure~\ref{#1}}
\newrefformat{tab}{Table~\ref{#1}}
\newrefformat{rmk}{Remark~\ref{#1}}
\newrefformat{clm}{Claim~\ref{#1}}
\newrefformat{def}{Definition~\ref{#1}}
\newrefformat{cor}{Corollary~\ref{#1}}
\newrefformat{lmm}{Lemma~\ref{#1}}
\newrefformat{prop}{Proposition~\ref{#1}}
\newrefformat{pr}{Proposition~\ref{#1}}
\newrefformat{app}{Appendix~\ref{#1}}
\newrefformat{prob}{Problem~\ref{#1}}
\newrefformat{ques}{Question~\ref{#1}}
\newrefformat{note}{Note~\ref{#1}}
\newrefformat{assump}{Assumption~\ref{#1}}
\newrefformat{issue}{Issue ~\ref{#1}}
\newrefformat{fix}{Fix ~\ref{#1}}

\newcommand{\reals}{\mathbb{R}}

\newcommand{\prob}[1]{\mathbb{P}\left(#1\right)}

\newcommand{\vect}[1]{\boldsymbol{#1}}

\newcommand{\bu}{\vect{u}}

\newcommand{\bx}{\vect{x}}
\newcommand{\by}{\vect{y}}

\newcommand{\balpha}{\vect{\alpha}}
\newcommand{\bbeta}{\vect{\beta}}

\newcommand{\lse}{\mathrm{LSE}}
\newcommand{\define}{\triangleq}

\newcommand{\pth}[1]{\left( #1 \right)}

\mathchardef\mhyphen="2D

\definecolor{MyGreen1}{RGB}{20,180,40}

\usepackage{siunitx}
\usepackage{color,fancyhdr}

\newcommand{\todo}{{\color{blue} TODO:}}

\usepackage[symbol]{footmisc}
\renewcommand{\thefootnote}{\fnsymbol{footnote}}

\newcommand{\tinyturbo}{\textsc{TinyTurbo }}
\newcommand{\tinyturbonosp}{\textsc{TinyTurbo}\ignorespaces}
\hypersetup{
pdftitle={A template for the arxiv style},
pdfsubject={q-bio.NC, q-bio.QM},
pdfauthor={David S.~Hippocampus, Elias D.~Striatum},
pdfkeywords={First keyword, Second keyword, More},
}

\begin{document}
\newcommand{\shorttitle}{}
\maketitle
\def\thefootnote{1}\footnotetext{University of Illinois, Urbana-Champaign}\def\thefootnote{2}\footnotetext{University of Texas, Austin}\def\thefootnote{\dag}\footnotetext{Equal contribution}
\def\thefootnote{*}
\blfootnote{Correspondance to: Ashwin, Rajesh <shebbar3@illinois.edu,  rajeshkmishra@utexas.edu>}
\begin{abstract}
In this paper, we introduce a neural-augmented 
decoder for Turbo codes called \tinyturbo. \tinyturbo has complexity comparable to the classical max-log-MAP algorithm but has  much better reliability than the max-log-MAP baseline and performs close to the MAP algorithm. We show that \tinyturbo exhibits strong robustness on a variety of practical channels of interest, such as EPA and EVA channels, which are included in the LTE standards. We also show that \tinyturbo strongly generalizes across different rate, blocklengths, and trellises. We verify the reliability and efficiency of \tinyturbo via  over-the-air experiments. 
\end{abstract}


\section{Introduction}
\label{sec:intro}

Codes form the backbone of the modern information age (WiFi, cellular, cable, and satellite modems). Some of the landmark codes include Reed-Muller (RM), BCH, Turbo, LDPC and Polar codes. In particular, Turbo codes have been widely used in modern communication systems and are part of 3G and 4G standards. While Turbo codes operate at near-optimal performance on the canonical additive white Gaussian noise (AWGN) channel, it is well known that the classical Turbo decoder \cite{berrou1993near} lacks robustness and performs poorly on non-AWGN channels. Thus designing turbo decoders with high reliability and robustness is of great interest. 

In recent years, we have had tremendous achievements of deep learning (DL) across various disciplines such as computer vision and natural language processing. In particular, we have also seen impressive results from DL based decoders in channel decoding \cite{nachmani2016,gruber2017deep,kim2018communication,kim2020physical,vasic2018learning, liang2018iterative, teng2019low, buchberger2020prunin, chen2021cyclically}. While these results demonstrate the potential of DL in designing optimal decoders, a majority of them still suffer from the huge computational complexity, which makes them intractable to be deployed in real-world communication systems.

Given this, more scalable approaches such as model-based DL has gained traction recently \cite{shlezinger2020model,qin2019deep}. One of the key ideas in this approach is to augment learnable parameters to existing channel decoders and train these parameters. A strong benefit of such neural decoders is that the complexity of the decoder does not increase much while reliability can be improved. In the context of Turbo codes, in a recent work ~\cite{turbonet}, He et al. introduce \textsc{TurboNet+}, which augments learnable parameters to decoding every bit, and show that it improves the reliability of the classical Turbo decoders. 

Despite the success of TurboNet+, there are three important open questions: (a) do we need all those learnable parameters, the number of which scales linearly in blocklength?  
(b) can we learn weights that generalize across rates, blocklengths, and codes and that are robust across channel variations? 
(c) what is the role of the learned weights? 
In this paper, we focus on addressing these questions.  
%
Our contributions are as follows. 
\begin{itemize}

    \item We propose \textsc{TinyTurbo}\footnote{Code can be found at https://github.com/hebbarashwin/tinyturbo}, a neural augmented Turbo decoder that has $18$ trainable weights.  
    We show that \textsc{TinyTurbo} recovers the reliability of the state-of-the-art neural augmented Turbo decoder, namely, TurboNet+~\cite{turbonet} which has $720$ weights, for AWGN channels. 
    \item We show the \emph{robustness} of \textsc{TinyTurbo}: \textsc{TinyTurbo} outperforms TurboNet+ and the classical Turbo decoder for several practical channels.
    \item We demonstrate the strong \emph{generalization} of \textsc{TinyTurbo}: \textsc{TinyTurbo} trained for rate-1/3 LTE Turbo codes of blocklength 40 performs well for Turbo codes with different blocklengths, rates, and trellises (e.g., blocklength 200, rate-1/2, Turbo-757). 

    
    

    \item Our over-the-air experiment demonstrates that \textsc{TinyTurbo} achieves improved reliability and efficiency compared to the classical Turbo decoders in indoor scenarios. 
    
    \item We provide the interpretation analysis, based on which we conjecture that the weights in \textsc{TinyTurbo} tend to mitigate the over-estimation of beliefs. 
\end{itemize}



\section{Background}
\label{sec:background}

\subsection{Turbo encoder}
\begin{figure}[!ht]
    \centering
    \includegraphics[width=\linewidth]{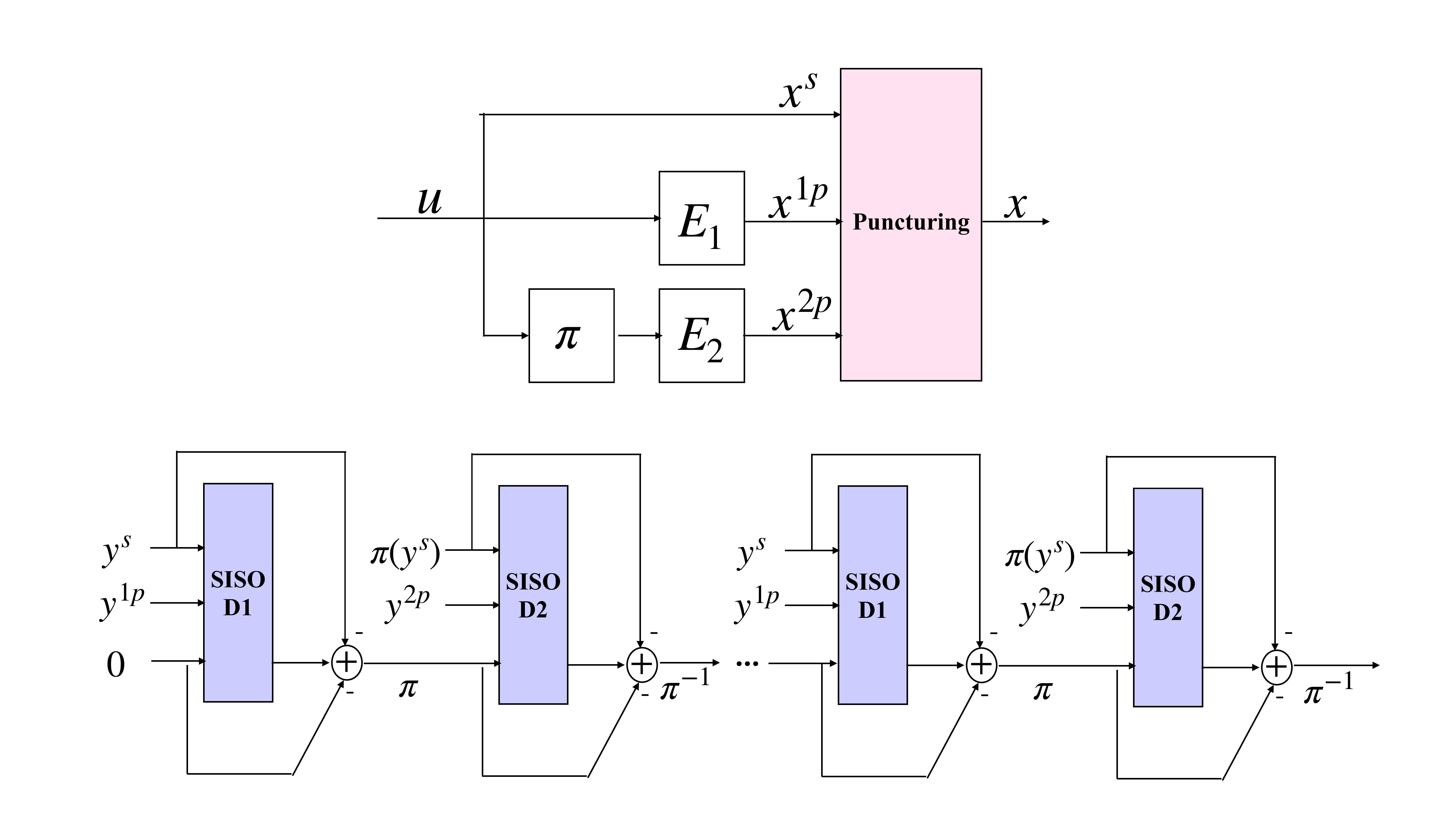}
    \caption{Turbo encoder (top); Turbo decoder (bottom)}
    \label{fig:turbo_encdecoder}
\end{figure}

Turbo encoder consists of an interleaver and two identical Recursive
Systematic Convolutional (RSC) encoders, denoted by $\pi$ and $(E_1, E_2)$ respectively, as depicted in Fig.~\ref{fig:turbo_encdecoder}. 
Let $\bu=(u_1,\ldots, u_K)$ denote the sequence of $K$ information bits that we wish to transmit. To encode the message bits $\bu$, a block of them is directly transmitted via the systematic bit sequence $\bx^s=(x^s_1,x^s_2,\ldots, x^s_K)$. On the other hand, the encoder $E_1$ generates the parity bit sequence $\bx^{1p}=(x^{1p}_1,x^{1p}_2,\ldots, x^{1p}_k)$ from $\bu$ whereas the encoder $E_2$ generates the parity sequence $\bx^{2p}=(x^{2p}_1,x^{2p}_2,\ldots, x^{2p}_k)$ using the interleaved input $\Tilde{\bu}=\pi(\bu)$. 
We assume that the encoded bits $\bx=(\bx^s, \bx^{1p}, \bx^{2p})$ are modulated via Binary Phase Shift Keying (BPSK) and transmitted over the channel. For concreteness, we consider LTE Turbo codes, for which the RSC code has the generator matrix $(1, g_1(D)/g_2(D))$ with $g_1(D)= 1+ D^2+ D^3$ and $g_2(D)= 1+D+D^3$~\cite{3gpp_lte}. 

\vspace{-1em}
\vspace{-1em}

\subsection{Turbo decoder}
As depicted in Fig.~\ref{fig:turbo_encdecoder} (bottom), the classical turbo decoder in \cite{berrou1993near} contains two identical Soft-Input Soft-Output (SISO) decoders $(D_1, D_2)$ and relies on the `Turbo principle', where these SISO decoders iteratively refine the posterior of the information bits. More precisely, each SISO decoding block takes the received signals and the prior as input and estimates the posterior distribution for the message bits, which is then fed as a prior for the next SISO decoding block. This procedure is repeated for a fixed number of iterations to estimate the final posterior. 
For the SISO decoding block, \cite{bahl1974optimal} uses the celebrated BCJR algorithm, which we explain below in the context of decoder $D_1$.


\emph{MAP turbo decoder:} Let $\by=(\by^s, \by^{1p})$ be the log-likelihood-ratios (LLRs) of the encoded bits corresponding to encoder $E_1$. The BCJR SISO decoder $D_1$ takes the soft-information $\by$ as its input and obtains the LLRs for the information bits as follows:
\begin{align*}
    L(u_k|\by) \define \log  \frac{\prob{u_k=1|\by}}{\prob{u_k=0|\by} }  = \log  \frac{\sum_{(s', s) \in S^1} \prob{s', s, \by}} {\sum_{(s', s) \in S^0} \prob{s', s, \by}} ,
\end{align*}
where $S^1=\{(s',s): u_k=1 \}$ denotes the set of all ordered pairs $(s',s)$ corresponding to state transitions $s' \rightarrow s$ caused by data input $u_k = 1$, whereas $S^0=\{(s',s): u_k=0 \}$ pertains to the input $u_k=0$. Assuming that the underlying channel is AWGN and memoryless, the joint probabilities $\prob{s', s, \by}$ can be efficiently computed:
\begin{align*}
    \prob{s', s, \by} &= \prob{s', s, \by_{1}^{k-1}, y_k, \by_{k+1}^K} \\
    & = \prob{s', \by_{1}^{k-1}} \prob{y_k, s|s'} \prob{\by_{k+1}^K |s} \\
    & = \alpha_{k-1}(s') \gamma_k(s',s) \beta_k(s),
\end{align*}
where $\alpha_{k-1}(s') \define \prob{s', \by_{1}^{k-1}}$ and $\beta_k(s) \define \prob{\by_{k+1}^K |s} $ can be computed via the forward and backward recursions \cite{bahl1974optimal}:
\begin{align*}
    \alpha_k(s)\!=\!\sum_{s' \in S_R}\!\alpha_{k-1}(s') \gamma_k(s',s), \\
    \beta_{k-1}(s')\!=\!\sum_{s \in S_R}\!\beta_k(s) \gamma_k(s',s)
\end{align*}
with the initial conditions $\alpha_0(s)=\beta_K(s)=\mathbbm{1}\{s=0\}$. Here $S_R =\{0,1,\ldots, 2^m-1 \}$ denotes the set of all possible states for the encoder whose memory is $m$. The branch transition probabilities $\gamma_k(s',s) \define \prob{y_k,s|s'}$ can be computed as
\begin{align}
    \gamma_k(s',s) = \mathrm{exp} \pth{\frac{1}{2}\pth{x_k^s y_k^s + x_k^{1p} y_k^{1p}} + \frac{1}2 u_k L(u_k)}, 
    \label{eq:branc_prob}
\end{align}
where $L(u_k)$ is the \emph{apriori} LLR for the bit $u_k$. Equivalently, by considering the logarithmic values of the above probabilities, i.e. $\bar{\alpha}_k(s) \define \log \alpha_k(s), \bar{\beta}_k(s) \define \log \beta_k(s)$, \newline and $ \bar{\gamma}_k(s',s) \define \log \gamma_k(s',s)$, we obtain
\begin{equation}
\begin{aligned}
   \bar{\gamma}_k(s',s) & = \frac{1}{2}\pth{x_k^s y_k^s + x_k^{1p} y_k^{1p}} + \frac{1}2 u_k L(u_k), \\
   \bar{\alpha}_k(s) &=  \lse_{s' \in S_R} \pth{\bar{\alpha}_{k-1}(s') + \bar{\gamma}_k(s',s) }, \\ 
   \bar{\beta}_{k-1}(s') &= \lse_{s \in S_R}\pth{ \bar{\beta}_k(s) + \bar{\gamma}_k(s',s) },\\
   L(u_k|\by) &= \lse_{(s',s) \in S^1} \pth{\bar{\alpha}_{k-1}(s') + \bar{\gamma}_k(s',s) + \bar{\beta}_k(s) }
  \\
  & \hspace{1em} - \lse_{(s',s) \in S^0} \pth{\bar{\alpha}_{k-1}(s') + \bar{\gamma}_k(s',s) + \bar{\beta}_k(s) }
   \end{aligned}
   \label{eq:lsequations}
\end{equation}
where $\lse(z_1,\ldots, z_n) \define \log (\exp(z_1)+\ldots+\exp(z_n))$.
Upon obtaining the posterior LLR $L(u_k|\by)$, the decoder $D_1$ computes the extrinsic LLR $L_e(u_k)$ as
\begin{align}
    L_e(u_k) = L(u_k|\by) - L(y_k^s) - L(u_k), \quad k \in [K]
    \label{eq:main_eq}
\end{align}
which is interleaved and passed as a prior to the decoder $D_2$.
The posterior $L^M(u_k|\by)$ after the $M^{th}$ iteration is used to estimate the message bits:
$$\hat{u_k} = \mathbbm{1}\{L^M(u_k|\by) < 0\}$$
\subsection{Max-log-MAP turbo decoder}
The \emph{MAP} algorithm uses the exact $\lse$ function in the above set of equations in Eq.~\eqref{eq:lsequations} which is computationally expensive. Hence in practice, several variants and approximations are often used. A popular such decoder is the \emph{max-log-MAP} algorithm. The main idea behind the max-log-MAP is to approximate the $\lse$ function by the maximum:
\begin{align*}
    \lse(z_1,\ldots, z_n) \approx \max(z_1,\ldots, z_n), \quad z_1,\ldots,z_n \in \reals.
\end{align*}
While the max-log-MAP algorithm is more efficient than the MAP, it is less reliable than the MAP \cite{ryan2009channel}. 


\subsection{Weighted Max-log-MAP turbo decoder} 

It is well-known in the literature \cite{vogt2000mlmap, chaikalis2002, yue2010, clau2003mlmap, sun2020mlmap} that the max-log-MAP max-log-MAP algorithm overestimates the LLRs of the message bits, which leads to propagation of errors in the later decoding stages. To counter this, a common technique is to scale the LLRs in Eq.~\eqref{eq:main_eq}. A variety of methods have been proposed to determine these scaling weights. Most existing works \cite{vogt2000mlmap, chaikalis2002, yue2010} propose using heuristics via off-line time-averaged estimates to obtain the best scaling weights for a specific SNR and channel setting. While Claussen et al. \cite{clau2003mlmap} propose to find the scaling weights that maximize the mutual information, heuristics are still needed as it is hard to estimate the mutual information. In \cite{sun2020mlmap}, Sun and Wang use brute force search to find these parameters, which is computationally prohibitive.

\section{\textsc{TinyTurbo}}
\label{sec:tinyturbo}

In the previous section, we saw the shortcomings of existing turbo decoders. Given this, it would be advantageous to have a set of scaling weights that are both robust across varying block lengths, codes, and channels. 
\emph{How do we find such weights?}

In a recent work, He et al. \cite{turbonet} introduce TurboNet+, which can be viewed as a weight-augmented version of the max-log-MAP algorithm. TurboNet+ augments the standard max-log-MAP algorithm by adding learnable weights for each bit index $k\in [K]$ in Eq.~\eqref{eq:main_eq}: $ L_e(u_k) = w^{(1)}_k L(u_k|\by) - w^{(2)}_k y_k^s - w^{(3)}_k L(u_k), \quad k \in [K]$.

\begin{figure*}[htbp]
  \centerline{\subfigure[Turbo$(40,132)$]{  \includegraphics[width=0.33\textwidth]{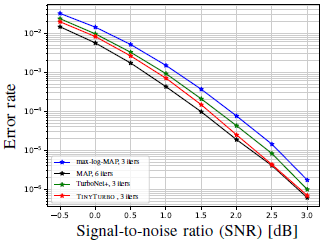}
    \label{fig:bl40_awgn}}
    \hfil

    \hfil

    \subfigure[Turbo$(200,412)$]{\includegraphics[width=0.33\textwidth]{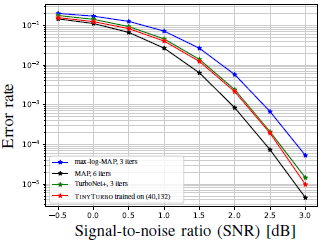}
        \label{fig:bl200_pun_awgn}}
        
    \subfigure[Turbo-$757$]{    \includegraphics[width=0.33\textwidth]{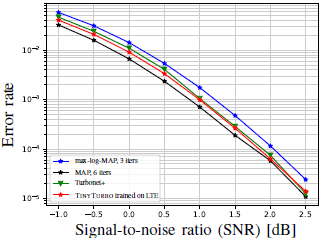}
    \label{fig:bl40_757}}
    }
    
  \caption{AWGN results: (a) \tinyturbo trained on Turbo $(40,132)$ outperforms the baselines and is close to MAP; (b) \& (c) \tinyturbo generalizes to different block lengths, code rates, and trellises.}
  \label{fig:awgn_results}
\end{figure*}

\begin{figure*}[!ht]
  \centerline{\subfigure[Bursty]{  \includegraphics[width=0.33\textwidth]{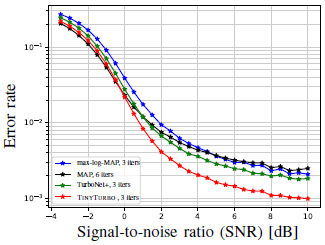}  \label{fig:bursty}}
    \hfil
    \subfigure[EPA]{\includegraphics[width=0.33\textwidth]{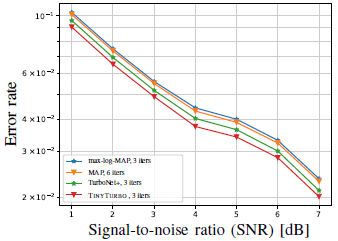}
    \label{fig:epa}}
    \hfil
    \subfigure[EVA]{\includegraphics[width=0.33\textwidth]{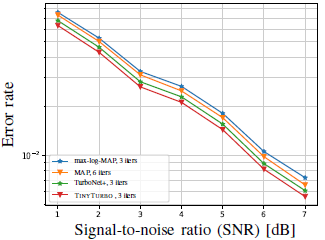}
    \label{fig:eva}}
    }
  \caption{Robustness: \tinyturbo exhibits strong robustness than baselines on practical channels: Bursty, EPA and EVA}
  \label{fig:robustness}
\end{figure*}

We propose \textsc{TinyTurbo}, a data-driven turbo decoder where the weights in the TurboNet+ architecture are entangled across bit positions. A major difference with TurboNet+ and other existing techniques is that the \textsc{TinyTurbo} efficiently learns the optimal set of weights directly from the data using an \emph{end-to-end loss function}. Learnt in such a purely data-driven manner, we show in Sec.~\ref{sec:results} (Fig.~\ref{fig:awgn_results} and Fig.~\ref{fig:robustness})
that these weights yield much better reliability results than the existing baselines and successfully generalize across various channels, block lengths, code rates, and code trellises. 

\begin{figure}[!ht]
    \centering
    \includegraphics[trim={0 10cm 0 6cm},clip, width=\linewidth]{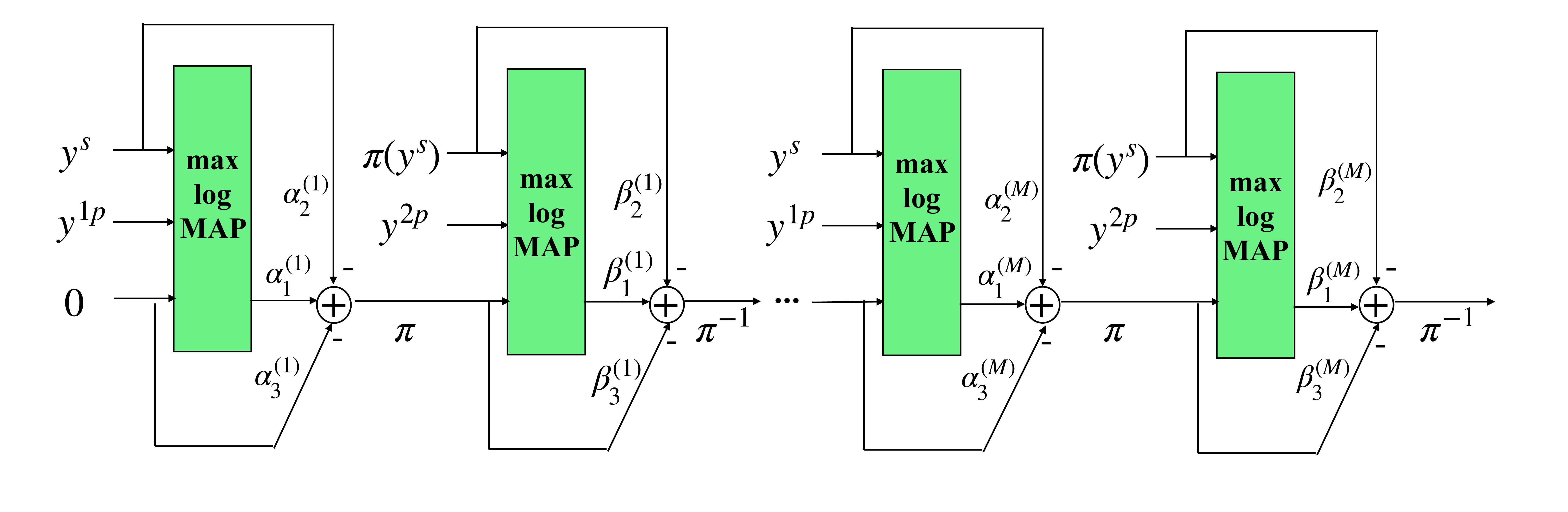}
    \caption{\textsc{TinyTurbo} decoder.}
    \label{fig:tiny_turbo}
\end{figure}

{\bf Architecture.} As depicted in Fig.~\ref{fig:tiny_turbo}, \textsc{TinyTurbo} generalizes the standard max-log-MAP algorithm by adding just \emph{three} trainable parameters $\balpha = (\alpha_1, \alpha_2, \alpha_3) \in \reals^3$ in the extrinsic LLR computation, i.e.
\begin{align}
    L_e(u_k) = \alpha_1  L(u_k|\by) - \alpha_2 L(y_k^s) - \alpha_3 L(u_k),  k \in [K].
    \label{eq:tinyturbo_eq}
\end{align} 
Note that the same parameters $\balpha$ are used for all the bit indices $k \in [K]$, making it amenable to generalize across block lengths. Similarly, decoder $D_2$ has three additional parameters $\bbeta = (\beta_1, \beta_2, \beta_3) \in \reals^3$. Thus every decoding iteration in \textsc{TinyTurbo} has $6$ parameters $(\balpha, \bbeta)$ with the total parameters being $6 M$, where $M$ denotes the number of decoding iterations. We consider $M=3$ in this paper. 

{\bf Training.} We propose an end-to-end loss function framework to train the \textsc{TinyTurbo} parameters $(\balpha_1^M, \bbeta_1^M)$. Hence these weights can be learnt directly from data alone. More precisely, let $\bu^{(1)}, \bu^{(2)}, \ldots, \bu^{(B)} \in \{0,1\}^K$ denote a batch of $B$ message blocks each of length $K$ and let $\by^{(1)}, \by^{(2)}, \ldots, \by^{(B)} \in \reals^N$ be the corresponding codeword-LLRs received by the \textsc{TinyTurbo} decoder. Let $L^M(\bu^{(i)}|\by^{(i)}) \in \reals^K$ denote the estimated LLRs of the message bits after $M$ decoding iterations of \textsc{TinyTurbo} for each block $i \in [B]$. Define the Binary Cross-Entropy (BCE) loss $L(\balpha_1^M, \bbeta_1^M)$ as
\begin{align*}
    L(\balpha_1^M, \bbeta_1^M)  &\define \frac{1}{B}  \sum_{i=1}^B \sum_{k=1}^K u^{(i)}_k \log  \sigma(L^M_k(\bu^{(i)}|\by^{(i)})) \\
    &\hspace{4em} + (1-u^{(i)}_k) \log \sigma(-L^M_k(\bu^{(i)}|\by^{(i)})).
\end{align*}
The weights ($\balpha_1^M, \bbeta_1^M$) are then trained by running stochastic gradient descent (SGD), or its variants like Adam \cite{kingma2017adam}, on the loss $L$.
The hyper-parameters for training are shown in Table~\ref{tab:hyperparams} below. Once trained, these weights are frozen and are used directly for inference (turbo decoding). 


\begin{table}[!htb]
\centering

\begin{tabular}{l|l}
Loss            & BCE                           \\
Optimizer       & Adam with initial LR = $0.0008$ \\
Batch size      & $1000$                          \\
Training SNR      & $-1$ dB                          \\
Number of iterations & $5000$                          \\ 
\end{tabular}
\caption{Training hyperparameters}
\label{tab:hyperparams}
\end{table}

{\bf Comparison with TurboNet+}.
While \textsc{TinyTurbo} is similar to TurboNet+, the following are the key differences: ($a$) TurboNet+ introduces weights for each bit index $k\in [K]$ in Eq.~\eqref{eq:main_eq}. 
As shown in Table~\ref{tab:num_params}, they have a total of $6MK$ parameters, as opposed to just $6M$ for \textsc{TinyTurbo}; ($b$) For training the weights, they consider a mean square error (MSE) between their estimated LLRs and the target BCJR-LLRs, which are obtained by running the MAP algorithm for BCJR. This introduces a computational overhead and makes the training slow as BCJR-LLRs are required for each training iteration; 
($c$) Since the weights are learned for each bit index $k \in [K]$, they cannot be reused for longer block lengths. In contrast, we show that by using just a $1/K$-fraction of weights and an end-to-end loss function, \textsc{TinyTurbo} learns a better set of weights that yield better reliability performance on various practical channels (Fig.~\ref{fig:robustness}). Further, our decoder generalizes better across a variety of scenarios (Fig.~\ref{fig:awgn_results}).

\begin{table}[!htb]
\centering
\begin{tabular}{l|ll}
Block length & 40  & 200  \\ \hline
\tinyturbo    & \textbf{18}  & \textbf{18}   \\
TurboNet+    & 720 & 3600
\end{tabular}
\caption{Number of parameters of \tinyturbo is independent of block length }
\label{tab:num_params}
\end{table}


\section{Results}
\label{sec:results}

We compare the performance of \tinyturbo with that of the standard max-log-MAP and MAP turbo decoders, and TurboNet+ \cite{he2020modeldriven}. We consider rate-$1/3$ and rate-$1/2$ Turbo-LTE codes of block length $40$ and $200$, and use the standard Quadratic Permutation Polynomial (QPP) interleaver \cite{3gpp_lte}. 
\subsection{AWGN channel}

We train the \tinyturbo for Turbo$(40,132)$ on the AWGN channel at $-1$ dB. Using these weights, we evaluate the performance of \tinyturbo on the following three codes under AWGN: Turbo$(40,132)$, Turbo-$757$, and Turbo$(200,412)$. As highlighted in Figs.~\ref{fig:bl40_awgn}, \ref{fig:bl200_pun_awgn}, and \ref{fig:bl40_757}, \tinyturbo with 3 decoder iterations achieves a bit error rate (BER) performance close to that of the MAP decoder with 6 iterations. Further, we achieve this with a decoding complexity comparable to a max-log-MAP decoder. We also note that \textsc{TinyTurbo} outperforms TurboNet+ even with $40\times$ fewer parameters. 

\subsection{Generalization to other blocklengths, rates, and trellises.}
\label{sec:block_len_gen} One of the major shortcomings of learning-based decoders is that training them directly on larger block lengths is generally difficult due to insufficient GPU memory, and unstable training \cite{jiang2019deepturbo, ebada2019deep, jiang2020learn, makkuva2021ko}. It is thus ideal to have models that are trained on short block lengths and are reusable for longer ones. Indeed, as illustrated in Fig.~\ref{fig:bl200_pun_awgn}, we observe that \textsc{TinyTurbo} trained on a small code, Turbo$(40,132)$, shows good performance when tested on a Turbo code of block length $200$. Interestingly, we also notice that a \tinyturbo model trained directly on block length $200$ achieves similar performance. These results highlight that our model seamlessly generalizes to longer block lengths. 

Puncturing is a technique used to encode and decode codes of higher rates using standard rate 1/3 encoders and decoders. Higher rates are achieved by removing certain parity bits according to a fixed pattern. The use of puncturing improves the flexibility of a system without significantly increasing its complexity. As shown in Figure \ref{fig:bl200_pun_awgn}, the weights learned using a rate $1/3$ Turbo code can be reused for a rate $1/2$ punctured Turbo code, which demonstrates the generalizability of \tinyturbo across code rates. 
%
%
Surprisingly, \textsc{TinyTurbo} also generalizes well to different trellises; Fig.~\ref{fig:bl40_757} shows that \textsc{TinyTurbo} trained on Turbo-LTE achieves a BER performance close to MAP even on a Turbo-$757$ code with the generator matrix $(1, \!1\!+\!D^2/\!1\!+\!D\!+\!D^2)$.


\subsection{Robustness across channel variations}
 \label{sec:robustness}

Here we evaluate the \textsc{TinyTurbo} decoder trained on AWGN on various non-AWGN settings (without any retraining). We consider both the simulated and practical channels.

We first test on a bursty channel, defined as $y = x+z+w$, where $z \sim \mathcal{N}(0, {\sigma_1}^2)$ is an AWGN noise, and $w \sim \mathcal{N}(0, {\sigma_b}^2)$ is a bursty noise with a high noise power and low probability of occurrence $\rho$. Here we consider $\sigma_b = 5$ and $\rho = 0.01$. Fig.~\ref{fig:bursty} highlights that \tinyturbo is significantly more robust to the bursty noise compared to the canonical Turbo decoders. 

We now demonstrate the robustness of \textsc{TinyTurbo} on various practical channels. In particular, we consider the multi-path fading channels, as defined in the 3GPP \cite{3gpp_channel}. We simulate them using the LTE toolbox in MATLAB. Specifically, we test on the Extended Pedestrian A model (EPA) and the Extended Vehicular A model (EVA) specified in the LTE standard.
The EPA and EVA channels represent a low and medium delay spread environment respectively. 
The received signals are equalized using linear MMSE estimation before proceeding to decode. As shown in Figures \ref{fig:epa} and \ref{fig:eva},  \textsc{TinyTurbo} achieves better reliability than the MAP Turbo decoder on the EPA and EVA channels. 

\section{Over-the-air} 
\label{sec:ota}

\begin{figure}[htbp]
  \centering
  \includegraphics[width=0.5\textwidth]{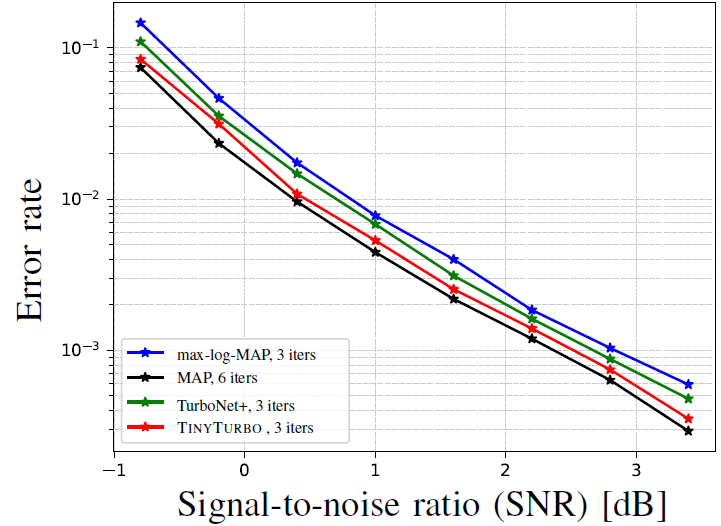}
  \label{fig:bl40_ota_robustness}
 \end{figure}

In this section, we use a rapid prototype system with a ZynQ SoC and an AD9361 transceiver as a testbed to evaluate the performance of our \tinyturbo algorithm in a practical setting. We implement these algorithms on the ZynQ CPU and the FPGA with an OFDM-based end-to-end communication system. Specifically, we analyze the robustness of our algorithm due to the multipath channel variations and showcase the practical viability of the algorithms on transceiver chips. In a real-time processing system, we use two of these systems as transmitter and receiver, to transmit and capture frames. The Turbo-encoded codewords were modulated and sent over the channel after adding a preamble for synchronization and frequency offset correction. The received data was equalized and demodulated to get LLR values for the decoder. Simulations were done over different gain settings at the transmitter and the receiver to get the results for different SNRs. As demonstrated in Figure ~\ref{fig:bl40_ota_robustness}, \tinyturbo is robust to multipath fading.

\section{Interpretation and Ablation Study}








\subsection{Interpreting \tinyturbo} 
 
\begin{figure}[htbp]
 \includegraphics[width=\textwidth]{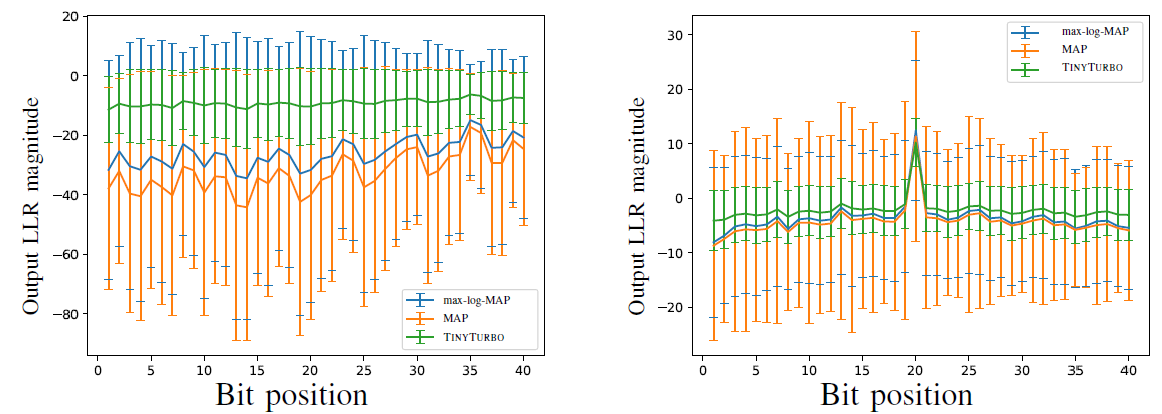}
    \caption{Average output LLRs on AWGN (left) and bursty (right) channels}
    \label{fig:out_llr}
\end{figure}

 In Sections~\ref{sec:results} and \ref{sec:ota}, we have demonstrated that \tinyturbo\footnote{The learned weights are shown in Table~\ref{tab:tinyutrbo}.} can achieve much better reliability than the baseline turbo decoders across a variety of channels. This raises a natural question: \emph{where do these gains come from?} To this end, we fix the input message bits to be the all-zero vector and examine the corresponding LLR values predicted by \tinyturbo and our baselines for Turbo(40,132). In Fig.~\ref{fig:out_llr} (left), we consider the AWGN channel and plot the mean LLR together with the error bars corresponding to two standard deviations. While all the decoding algorithms have a negative mean LLR since the message bits are all-zero, only \tinyturbo and MAP decoders can keep the deviations close to/lesser than zero resulting in fewer errors. On the other hand, max-log-MAP has a significant portion of zero crossings and hence more decoding errors. These observations are consistent with our AWGN results in Fig. ~\ref{fig:bl40_awgn}. 

We consider a simplified bursty channel model $y = x+z+w$, where $x, y, z$ are as described in Section \ref{sec:robustness}, and $w = 10.0$ only at the $57^{th}$ symbol in the 132-length zero codeword. Figure \ref{fig:out_llr} (right) illustrates that only \tinyturbo can keep the deviations contained below zero, whereas the baselines have lots of zero crossings resulting in poor decoding performance. This explains the superior performance of \tinyturbo as demonstrated in Fig.~\ref{fig:bursty}.

\begin{table}[h]
\centering
\resizebox{0.5\columnwidth}{!}{
\begin{tabular}{llll|lll}
                                 & $\alpha_1$     & $\alpha_2$     & $\alpha_3$    & $\beta_1$     & $\beta_2$    & $\beta_3$    \\ \hline
\multicolumn{1}{l|}{Iteration 1} & 0.445  & 0.584  & 1     & 0.641  & 0.779 & 0.662 \\
\multicolumn{1}{l|}{Iteration 2} & 0.834  & 0.795  & 0.725 & 0.863  & 0.716 & 0.645 \\
\multicolumn{1}{l|}{Iteration 3} & 0.911  & 0.715  & 0.638 & 0.263  & 0.616 & 0.938
\end{tabular}}
\caption{\tinyturbo weights}
\label{tab:tinyutrbo}
\end{table}

\subsection{Ablation study}

\begin{figure}[htbp]
  \centering
  \includegraphics[width=0.5\textwidth]{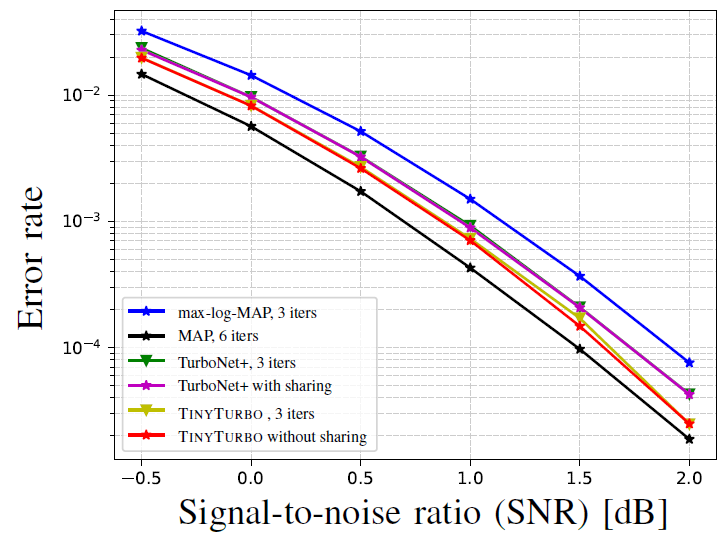}
 \caption{Turbo$(40,132)$. Our gains are mainly due to the end-to-end BCE loss function.}
  \label{fig:bl40_ablation}

\end{figure}

The key differences between \textsc{TinyTurbo} and TurboNet+ are: (i) weights are shared across bit positions for \tinyturbo unlike TurboNet+, and (ii) \tinyturbo minimizes the BCE loss between the predicted LLRs and the ground truth message bits; TurboNet+ minimizes the MSE loss between the estimated LLRs and the BCJR-LLRs from the MAP algorithm.
We evaluate the contributions of these components to our gains using the following ablation experiments: (i)  we do not entangle weights across bit positions in \tinyturbo but train them using BCE loss. The resulting red curve in \prettyref{fig:bl40_ablation} highlights that this approach has the same performance as that of \tinyturbonosp. (ii) we entangle weights but train them using the same procedure as TurboNet+. The corresponding purple curve in \prettyref{fig:bl40_ablation} shows similar performance as TurboNet+. Together, these experiments suggest that training the weights via the end-to-end BCE loss function approach is the key contributing factor to the gains of \tinyturbo over TurboNet+. While sharing weights did not yield any change in the performance, it nonetheless allows for a more computationally efficient decoder with a low-memory requirement.

\section{Acknowledgement}
This work was supported by ONR grants W911NF-18-1-0332, N00014-21-1-2379, NSF grant CNS-2002932, Intel, and the affiliates of the 6G@UT center within WNCG at UT Austin.
\newpage

\bibliographystyle{IEEEtran}
\bibliography{references_commu}

\end{document}